\documentclass[12pt]{article}
\input{epsf.sty}

\def\fo{\hbox{{1}\kern-.25em\hbox{l}}}

\def\slashchar#1{\setbox0=\hbox{$#1$}           % set a box for #1
   \dimen0=\wd0                                 % and get its size 
   \setbox1=\hbox{/} \dimen1=\wd1               % get size of /
   \ifdim\dimen0>\dimen1                        % #1 is bigger 
      \rlap{\hbox to \dimen0{\hfil/\hfil}}      % so center / in box 
      #1                                        % and print #1
   \else                                        % / is bigger 
      \rlap{\hbox to \dimen1{\hfil$#1$\hfil}}   % so center #1
      /                                         % and print /
   \fi}                                         %

\def\ppar{p_\parallel}
\def\pperp{p_\perp}
\def\hide#1{[hidden stuff]}
\def\bibitemx#1{\bibitem{#1}}
\def\titre#1{{\it ``#1,''}}

\def\beq{\begin{equation}}
\def\eeq{\end{equation}}
\def\eq{\end{equation}}
\def\to{\rightarrow}

\def\mEt{\mbox{${\hbox{$E$\kern-0.6em\lower-.1ex\hbox{/}}}_T$}\, } %missing ET

\def\bsg{\ifmmode B\to X_s\gamma\else $B\to X_s\gamma$\fi}
\def\bsll{\ifmmode B\to X_s\ell^+\ell^-\else $B\to X_s\ell^+\ell^-$\fi}
\def\bstt{\ifmmode B\to X_s\tau^+\tau^-\else $B\to X_s\tau^+\tau^-$\fi}
\def\shat{\ifmmode \hat{s}\else $\hat{s}$\fi}

\newcommand{\newc}{\newcommand}

\newc{\lcal}{\int {\cal L}dt}
 
\newc{\LSP}{{\chi^0_1}}
\newc{\stauR}{{\tilde \tau_R}}
\newc{\stau}{{\tilde \tau_1}}
\newc{\mstop}{m_{\tilde{t}}}
\newc{\mHpm}{m_{H^\pm}}
\newc{\gsim}{\lower.7ex\hbox{$\;\stackrel{\textstyle>}{\sim}\;$}}
\newc{\lsim}{\lower.7ex\hbox{$\;\stackrel{\textstyle<}{\sim}\;$}}
\newc{\ie}{{\it i.e.}}          
\newc{\etal}{{\it et al.}}
\newc{\eg}{{\it e.g.}}          
\newc{\kev}{\hbox{\rm\,keV}}            
\newc{\mev}{\hbox{\rm\,MeV}}            
\newc{\gev}{\hbox{\rm\,GeV}}            
\newc{\tev}{\hbox{\rm\,TeV}}
\newc{\xpb}{\hbox{\rm\, pb}}
\newc{\xfb}{\hbox{\rm\, fb}}

\newc{\mtop}{m_t}
\newc{\mbot}{m_b}
\newc{\mz}{m_Z}
\newc{\mw}{M_W}
\newc{\alphasmz}{\alpha_s(m_Z^2)}
\newc{\swsq}{\sin^2\theta_W}
\newc{\tw}{\tan\theta_W}
\newc{\cw}{\cos\theta_W}
\newc{\sw}{\sin\theta_W}
\newc{\BR}{\hbox{\rm BR}}
\newc{\zbb}{Z\to b\bar}
\newc{\Gb}{\Gamma (Z\to b\bar b)}
\newc{\Gh}{\Gamma (Z\to \hbox{\rm hadrons})}
\newc{\rbsm}{R_b^\hbox{\rm sm}}
\newc{\rbsusy}{R_b^\hbox{\rm susy}}
\newc{\drb}{\delta R_b}

\newc{\sgn}{\mbox{sgn}}
\newc{\tbeta}{\tan\beta}
\newc{\uL}{{\tilde u_L}}
\newc{\uR}{{\tilde u_R}}
\newc{\cL}{{\tilde c_L}}
\newc{\cR}{{\tilde c_R}}
\newc{\tL}{{\tilde t_L}}
\newc{\tR}{{\tilde t_R}}
\newc{\dL}{{\tilde d_L}}
\newc{\dR}{{\tilde d_R}}
\newc{\sL}{{\tilde s_L}}
\newc{\sR}{{\tilde s_R}}
\newc{\bL}{{\tilde b_L}}
\newc{\bR}{{\tilde b_R}}
\newc{\eL}{{\tilde e_L}}
\newc{\eR}{{\tilde e_R}}
\newc{\mhp}{m_{H^\pm}}
\newc{\mhalf}{m_{1/2}}
\newc{\emt}{{e/\mu /\tau}}

\newc{\lR}{\tilde{l}_R}
\newc{\lL}{\tilde{l}_L}
\newc{\nL}{\tilde{\nu}_L}
\newc{\na}{\chi^0_1}
\newc{\nb}{\chi^0_2}
\newc{\nc}{\chi^0_3}
\newc{\nd}{\chi^0_4}
\newc{\ca}{\chi^{\pm}_1}
\newc{\cb}{\chi^{\pm}_2}
\newc{\camp}{\chi^\mp_1}
\newc{\cbmp}{\chi^\mp_1}
\newc{\capos}{\chi^{+}_1}
\newc{\caneg}{\chi^{-}_1}
\newc{\phit}{\phi_t}
\newc{\phib}{\varphi_b}
\newc{\phiew}{\phi_{ew}}
\newc{\htz}{h^0_t}
\newc{\hbz}{h^0_b}
\newc{\hewz}{h^0_{ew}}
\newc{\hsmz}{h^0_{sm}}
\newc{\huz}{h^0_u}
\newc{\hsusyz}{h^0_{susy}}

\def\NPB#1#2#3{Nucl. Phys. B {\bf #1}, #3 (19#2)}
\def\PLB#1#2#3{Phys. Lett. B {\bf #1}, #3 (19#2)}

\def\PRD#1#2#3{Phys. Rev. D {\bf #1}, #3 (19#2)}
\def\PRL#1#2#3{Phys. Rev. Lett. {\bf#1}, #3 (19#2)}

\def\beq{\begin{equation}}
\def\eeq{\end{equation}}
\def\bea{\begin{eqnarray}}
\def\eea{\end{eqnarray}}
\catcode`@=11
\long\def\@caption#1[#2]#3{\par\addcontentsline{\csname
  ext@#1\endcsname}{#1}{\protect\numberline{\csname
  the#1\endcsname}{\ignorespaces #2}}\begingroup
    \small
    \@parboxrestore
    \@makecaption{\csname fnum@#1\endcsname}{\ignorespaces #3}\par
  \endgroup}
\catcode`@=12

\def\jfig#1#2#3{
 \begin{figure}
 \centering
 \epsfysize=3.0in
 \hspace*{0in}
 \epsffile{#2}
 \caption{#3}
 \label{#1}
 \end{figure}}
\begin{document}
%Remember: There Is No Cabal.
\begin{titlepage}

\begin{flushright}
%DRAFT \today \\
hep-ph/9906234 \\
SLAC-PUB-8119 \\
CERN-TH/99-139
\end{flushright}

%\vspace{1cm}

%\vspace*{17.3cm}

%\begin{flushleft}
%SLAC-PUB-XXXX \\
%CERN-TH/99-XX \\
%April 1999\\
%\end{flushleft}

%\vspace*{-19.0cm}
%Stamp out redundancy and avoid repetition.

%\vspace{.3in}
%\begin{center}

%\end{center}

\huge
%\vspace{0.05in}
%\renewcommand{\thefootnote}{\fnsymbol{footnote}}
\bigskip
\bigskip
\begin{center}
{\large\bf
Electroweak Precision Measurements and Collider Probes \\
of the Standard Model with Large Extra Dimensions}
\end{center}
%\vspace{0.2in}

\large

\vspace{.15in}
\begin{center}

Thomas G.~Rizzo$^a$\footnote{Work supported by the Department
of Energy under contract DE-AC03-76SF00515} and James D.~Wells$^b$

\small

\vspace{.1in}
{\it $^{(a)}$Stanford Linear Accelerator Center, Stanford, CA 94309 USA \\}
\vspace{0.1cm}
{\it $^{(b)}$CERN, Theory Division, CH-1211 Geneva 23, Switzerland \\}

\end{center}
 
%\vspace{0.2in}
 
\vspace{0.15in}
 
\begin{abstract}

The elementary particles of the Standard Model may live in more than
$3+1$ dimensions.  We study the consequences of large compactified
dimensions on scattering and decay observables at high-energy
colliders.  Our analysis includes global fits to electroweak precision
data, indirect tests at high-energy electron-positron colliders (LEP2
and NLC), and direct probes of the Kaluza-Klein resonances at hadron
colliders (Tevatron and LHC).  The present limits depend sensitively
on the Higgs sector, both the mass of the Higgs boson and how many
dimensions it feels. If the Higgs boson is trapped on a $3+1$
dimensional wall with the fermions, large Higgs masses (up to
$500\gev$) and relatively light Kaluza-Klein mass scales (less than
$4\tev$) can provide a good fit to precision data.  That is, a light Higgs
boson is not necessary to fit the electroweak precision data, as it is
in the Standard Model.  If the Higgs boson propagates in higher
dimensions, precision data prefer a light Higgs boson (less than 260
GeV), and a higher compactification scale (greater than 3.8 TeV).
Future colliders can probe much larger scales.  For example, a 1.5 TeV
electron-positron linear collider can indirectly discover Kaluza-Klein
excitations up to 31 TeV if $500\xfb^{-1}$ integrated luminosity is
obtained.

\end{abstract}

\medskip

\begin{flushleft}
June 1999
\end{flushleft}

\end{titlepage}

\baselineskip=18pt
\setcounter{footnote}{1}
\setcounter{page}{2}
\setcounter{figure}{0}
\setcounter{table}{0}
\tableofcontents
%\newpage

%\vfill
%\eject

%

%%%%%%%%%%%%%%%%%%%%%%%%%%%%%%%%%%%%%%%%%%%%%%%%%%%%%%%%%%%%%%%%%%%%%5
\section{Introduction}
\medskip

The original motivation for adding a large compact dimension was to generate
a $3+1$ dimensional vector 
gauge fields from a purely gravitational action in higher 
dimensions (see~\cite{appelquist87,oraifeartaigh98} for a review).
Describing nature completely by this mechanism is not viable.
Not only is matter unexplainable in this approach, but the $3+1$ dimensional
action inescapably contains a massless scalar particle that successfully
competes with a spin-2 particle (graviton) to create a strong
mix of scalar-tensor gravity unacceptable to modern experiment.
 
One conceptual cousin of the original Kaluza-Klein 
idea is 
string theory, or M theory, where strings and $D$-branes populate the
higher dimensional space rather than just a spin-2 
graviton (see~\cite{green87,polchinski98} for reviews).  
A strong motivation for string theory is that it may be finite,
and may thus provide a self-consistent description of quantum gravity.
String theory also predicts troublesome scalar moduli particles that
make it a challenge to identify the ground state of the theory. 
Solutions to this problem have been postulated, and progress has
been made on other aspects of the theory, giving hope that it may be
possible in time 
to write down a string theory description of nature.

Recently, it has been pointed out that there are more reasons to
suggest extra dimensions than just having a self-consistent description
of gravity~\cite{arkani-hamed98,arkani-hamed98b,antoniadis98a}.  
The additional motivations include new directions to
attack the hierarchy 
problem~\cite{arkani-hamed98}
and the cosmological
constant problem~\cite{sundrum97a,arkani-hamed98c}, 
unifying the gravitational coupling with the gauge 
couplings~\cite{witten96,horava95a,horava96a},
perturbative supersymmetry breaking in
string theory~\cite{kounnas88a,ferrara89}, and low-scale
compactifications of string theory~\cite{antoniadis90,lykken96a,
antoniadis98a,shiu98a,kakushadze98a}.  
An important breakthrough was the realization
that the gravitational scale could be as low as the weak scale and
still be phenomenologically viable~\cite{arkani-hamed98,arkani-hamed98b}. 
Two or more large extra dimensions
felt by gravity are required to make this possible. 
Another tantalizing realization is
that gauge couplings may unify with a greatly
reduced string scale if gauge fields feel one or more large extra 
dimensions~\cite{dienes98a}-\cite{kak99}.
A tentative picture is filling in for a 
viable scenario with TeV-scale extra dimensions, and especially
TeV-scale string theory with a vastly reduced Planck scale, compactification
scale, string scale, and unification scale.

In this article, we focus on the phenomenology of the
gauge and matter sectors of theories with large extra dimensions. 
In particular, the
Kaluza-Klein states of the gauge particles and matter particles can
have important observable consequences at high-energy colliders.
It is these consequences that we wish to study.  We build on previous
studies that assumed a similar framework and discussed relevant
collider phenomenology~\cite{antoniadis93a}-\cite{nath99c}.

In principle,
gravitational radiation into extra dimensions and virtual graviton induced 
observables are correlated with observables generated by KK excitations
of the gauge and matter fields.  Many detailed studies
on gravitational effects at high-energy 
colliders~\cite{giudice98a}-\cite{han99a}
and important astrophysical bounds~\cite{davidson,cullen99a,hall99b,barger99} 
have appeared.
However, to know the correlations between these effects and what we
study here requires that we either specify the underlying theory, or
assume that gravitational effects do not pollute the signals.  We choose
the latter path by assuming that gravity propagates in significantly more extra
dimensions than the gauge and matter fields do.  For example, gravity may
propagate in ten dimensions, while gauge fields are confined to a
$\ppar$-brane (gauge bosons) or 3-brane (fermions). 
($\ppar$ is defined to be the number of spatial dimensions
that bulk gauge fields feel, and $\pperp$ is defined to be the number of
remaining spatial dimensions in which gravity propagates.)
Also, we assume that the
higher-dimensional gravity scale and the gauge-unification scale
are comparable, as is expected in string theory.  These assumptions
imply that gravitational radiation will not be as significant as
gauge KK excitations in collider phenomenology. The exact strength of 
virtual graviton
exchange effects is not calculable, and so it is difficult to
tell how probing they are with respect to the gauge 
interactions pursued here.  Estimates based on naive dimensional analysis
suggest that the virtual graviton exchange processes
in some cases may be comparable
in probing power of extra dimensions as KK excitations of gauge bosons
given the above assumptions.

In the following sections we define a five-dimensional Standard Model
(5DSM).
Particularly important is the definition of the Higgs boson fields 
in this Lagrangian, since electroweak symmetry breaking effects will 
correlate strongly with some observables.
We then compactify the extra dimensions and work in an effective field
theory that is the Standard Model plus additional non-renormalizable
interactions arising from integrating out KK excitations. 
We then do a global fit to precision electroweak data and find limits
on the gauge compacification scale. Several comparisons of precision
electroweak data to the SM with extra dimensions 
have been published recently~\cite{nath99a,masip99a,nath99b,hall99a}.  
Our contributions in this direction are to construct a global fit  to
all relevant data, and to present results in terms of operator coefficients
rather than just a fifth dimension compactification scale. One result
from the global fit demonstrates that a light Higgs boson is not necessary,
in contrast to the Standard Model fit which requires it.
We also study the possibility
of finding the first excited state at hadron colliders, and derive
sensitivities to the full KK tower at $e^+e^-$ colliders.
In the last section we conclude and summarize the results.

%%%%%%%%%%%%%%%%%%%%%%%%%%%%%%%%%%%%%%%%%%%%%%%%%%%%%%%%%%%%%%
\section{The Standard Model in extra dimensions}
\bigskip

We begin by considering only one extra spatial dimension beyond the
usual $3+1$ dimensions.  Our first task is to state
which Standard Model particles live in five spacetime dimensions
and which live only in the four dimensions.  
In order to obtain massive chiral fermions we assume that the
fermions live in the ``twisted sector'' of string theory, and so
are naturally confined to ``walls'' of an orbifold fixed point
in the higher dimensional space.  The gauge fields are non-chiral and so may
live with impunity in higher dimensions, that is the fifth dimension,
or the ``bulk.''  These assumptions are 
essentially identical to those made in ref.~\cite{pomarol98a,delgado98a,
masip99a}.

%%%%%%%%%%%%%%%%%%%%%%%%%%%%%%%%%%%%
\subsection{The Higgs sector}

It is somewhat more difficult to decide what to do with the Higgs fields.
They are non-chiral fields as well, and with no reference to a more 
fundamental theory it appears natural to put them in the bulk with
the gauge fields.  To answer this question more satisfactorily, it is 
necessary to discuss the role of supersymmetry~\cite{pomarol98a,
antoniadis98b,delgado98a}.
The more fundamental theory is likely
to contain space-time supersymmetry.  Indeed, one of the motivations for
large extra dimensions is the ability to obtain tree-level supersymmetry
breaking at $R^{-1}$ from Scherk-Schwartz compactification of a
TeV string theory.  The superpartners will then have masses
near $R^{-1}$ and will have little effect on current collider
phenomenology as long as $R^{-1}\gsim 200\gev$.  

As a consequence 
of supersymmetry, two Higgs doublets are necessary in the spectrum,
$H_u$ which gives mass to up-type quarks, and $H_d$ which gives mass
to down-type quarks and leptons.
If one Higgs boson is
on the wall and the other Higgs boson is in the bulk, then successful
gauge coupling unification is possible with only the states of the
MSSM in the low-energy spectrum~\cite{dienes98b}.  Unification is also possible
by putting both Higgs fields in the bulk along with extra particles
that may be necessary for proton stability and other 
reasons~\cite{dienes98a,dienes98b,
kakushadze98b,kakushadze98c,kakushadze99a}.  Alternatively,
it may not be necessary to require both Higgs fields to be
zero-mode excitations under orbifolding~\cite{munoz,dienes98b}.

We therefore allow our Higgs sector to contain Higgs field(s)
in the bulk and Higgs field(s) on the wall~\cite{masip99a}.  We define
\beq
\tan\phi \equiv \frac{\langle \varphi_{\rm wall}\rangle}
       {\langle \varphi_{\rm bulk}\rangle},
\eeq
where $\langle \varphi_{\rm wall}\rangle$ is the vacuum expectation
value of the Higgs field on the wall, and $\langle \varphi_{\rm bulk}\rangle$
is the vacuum expectation
value of the Higgs field in the bulk.
In some theories $\varphi_{\rm wall}$ can be identified with either
$H_u$ or $H_d$, and $\varphi_{\rm bulk}$ can be identified with the
other Higgs field of the MSSM.  In these cases, $\tan\phi=\tan\beta$
or $\tan\phi =1/\tan\beta$, where $\tan\beta\equiv \langle H_u\rangle
/\langle H_d\rangle$.  However, the low-energy effective
theory may more natural best be described in terms of a single Higgs boson
originating from non-chiral bulk field(s), in which case
$\tan\phi =0$ ($\sin\phi =0$).
Furthermore, although supersymmetry may be necessary for a viable 
string scenario,
the most economical model is the Standard Model with one Higgs field
either in the $\ppar$-brane bulk or confined on the wall.
Therefore, the choices $\tan\phi=0$ and $\tan\phi=\infty$ will be of
particular interest when we discuss the EW precision measurement
predictions below.

%%%%%%%%%%%%%%%%%%%%%%%%%%%%%%%%%%%%%%%%%%%%%%%%%%%%%%%%%%%%
\subsection{The 5DSM Lagrangian and renormalized parameters}

Our starting framework is equivalent to ref.~\cite{pomarol98a,delgado98a,
masip99a}, where we assume the vector bosons and one Higgs field ($\varphi_b$)
live in the 5d bulk, and the fermions and another Higgs field ($\varphi_w$)
live on the 4d wall or boundary of the $S^1/Z_2$ orbifold.  
In five dimensions, the kinetic terms of the Lagrangian are simply
\beq
L_5 =  \int dy d^4x \left[
  -\frac{1}{4}F_{AB}^2+|D_A \varphi_b |^2+\left(
 i\bar \psi \sigma^\mu D_\mu \psi +|D_\mu \varphi_w|^2\right)\delta(y)
   +\ldots \right],
\eeq
where $\hat g$ is the $5d$ gauge coupling in the covariant derivative, 
and $\varphi_b$ is
the Higgs boson in the bulk, and $\varphi_w$ is the Higgs boson
on the wall. The $\delta(y)$ function indicates that the fermions
and $\varphi_w$ fields are localized at $y=0$, the location of
the 3-brane wall. 

Compactifying the fifth dimension on a $S^1/Z_2$
line segment, one finds
\bea
\label{4d Lagrangian}
L_4 & = & \int d^4x \left\{ \sum_{n=0}^\infty \left[ 
 -\frac{1}{4} F_{\mu\nu}^{(n)2} +\frac{1}{2}\left( \frac{n^2}{R^2}
  +2g^2|\varphi_b|^2\right) V_\mu^{(n)}V^{(n)\mu}\right] \right. \\
 & + & \left. 
   g^2 |\varphi_w|^2\left( V_\mu^{(0)}+\sqrt{2}\sum_{n=1}^\infty V_\mu^{(n)}
    \right)^2+i\bar\psi \sigma^\mu \left[ \partial_\mu
  +igV_{\mu}^{(0)}+ig\sqrt{2}\sum_{n=1}^\infty V_\mu^{(n)}\right] \psi
  +\ldots \right\}, \nonumber
\eea
where $g=\hat g/\sqrt{\pi R}$ is the four dimensional gauge coupling.
In the non-Abelian case, one should replace $V^{(n)}_\mu$ with
$\lambda^aV^{a(n)}_\mu$, where $\lambda^a$ are the group generators, 
to obtain the appropriate expressions.
From this Lagrangian interactions in the theory are specified.
The KK states have an additional $\sqrt{2}$ strength in their interactions,
which may appear odd at first sight.  This factor arises from 
rescaling the gauge kinetic terms to be canonically normalized
for all $n$.  Also, the zero-mode scalars from $V_{M>4}$ are not present
since $V_{M>4}$ fields are chosen to be odd under the $Z_2$ orbifolding.

Many of the renormalized coupling
parameters, such as the gauge couplings, 
of the 5DSM are directly analogous to the SM
parameters.  However, we emphasize that it is a
{\em different theory}. Even though these gauge
couplings ``look the same'' as the SM, they do not
relate the same way to physical observables measured
at high-energy colliders.  For this reason it is
more appropriate to ignore the Standard Model and
construct predictions for observables from our 5DSM Lagrangian
and compare to experiment.  These observables will depend
on gauge couplings, the compactification scale $M_c\equiv R^{-1}$,
and $\tan\phi$.

%%%%%%%%%%%%%%%%%%%%%%%%%%%%%%%%%%%%%%%%%%%%%%%%%%%%%%%%%%%%%%%%%
\subsection{The applicability of effective field theory}
\label{applicability}

A precise description
of the phenomenology requires a complete understanding of the underlying
theory.  This is especially true with two or more extra dimensions, since
the coefficients of operators induced by KK excitations are divergent
when trying to apply a naive effective field theory approach to integrating
out these modes.  More precisely, there is no theoretical problem with
constructing an effective
field theory description of low energies below the compactification scale,
and utilizing it to calculate all observables.  The difficulty is that there
is no model independent way to match all the couplings with the full theory.
The simplest approaches of compactifying field theories of higher dimensions to
field theories of lower dimensions often do not yield sensible results
for the effective theory.  

Specifically, in the effective theory there will be operators arising
from integrating out all the higher modes.  These operators will 
have coefficients that depend on
\beq
{\cal O}\sim \sum_{\vec n} \frac{g^2_{\vec n}}{\vec n\cdot \vec n}.
\eeq
For one extra dimension, 
\beq
\sum_{\vec n} \frac{g^2_{\vec n}}{\vec n\cdot \vec n} =\frac{g^2\pi^2}{6}
  ~~~({\rm assuming}~g_{\vec n}=g)
\eeq
which is convergent.  For two or more extra dimensions 
the sum diverges.
However, a more accurate application of the fundamental theory indicates
that $g$ depends on $\vec n$, and is in general given 
by~\cite{antoniadis94a,antoniadis99a}
\beq
\label{gn}
g_{\vec n} \sim g \, {\rm exp} 
   \left( \frac{-\vec n\cdot\vec n}{R^2 M_s^2} \right),
\eeq
where $M_s$ is the string scale.  This behavior is in qualitative
agreement with string scattering amplitudes at high energy which
tend toward zero.  The exponential suppression then cures the problem
of divergent summations of KK states.  However, the precise coefficients
and form of Eq.~\ref{gn} is model dependent.

Also, there are many other model dependent considerations that will yield
different couplings of KK gauge bosons to different fermions.  For example,
in ref.~\cite{arkani-hamed99a} it was pointed out that this situation
arises if fermions are
stuck to different points of a thick wall. In this case, the KK phenomenology
could be qualitatively different than what is presented here.  

In an effort to be as model independent as possible, we present all our
``indirect'' search results in terms of a parameter $V$ which is defined
to be
\beq
V\equiv 2\sum_{\vec n} \left( \frac{g^2_{\vec n}}{g^2}\right)
   \frac{m_W^2}{\vec n^2 M^2_c}.
\eeq
It is this quantity that can account for variations of $g$ for different
$\vec n$ in the summation of the correct effective theory, and the 
regularization of the KK sum. 
Often, for
concreteness, we will translate a limit of $V$ that we find into a limit
on $M_c$ by assuming one extra dimension and that $g_{\vec n}=g$ 
for all $\vec n$.
We must also keep in mind that other subtleties of the full theory
may contribute to collider phenomenology in addition to what we have
discussed here~\cite{shiu99a,bachas98,antoniadis98c}.

%%%%%%%%%%%%%%%%%%%%%%%%%%%%%%%%%%%%%%%%%%%%%%%%%%%%%%%%%%%%%%%%%%
\section{Precision measurements}
\bigskip

In  the Standard Model, all physical observables can be predicted in terms of a
small set of input observables. Equivalently, the Standard Model contains
several parameters in the Lagrangian which can be fit to by comparing
calculations within the model to measurements.  There are more 
observables than there are parameters, and so the fit is over-constrained.
A global $\chi^2$ analysis to precision electroweak data can determine
if a particular model, such as the SM, is a consistent description of nature.

In the following we do a global analysis of EW precision measurement data 
using the higher dimensional Standard Model (HDSM).  In the limit
that the extra compactified dimensions' radii tend to zero, we will
recover the Standard Model global fit results.  It has been often stated
that the SM fits the EW precision data very well; however, this is only
true if we assume that the Higgs boson is light.  In the 5DSM there
are two more parameters in the theory beyond the usual SM parameters
that will impact precision
measurement predictions~\cite{nath99a,masip99a,nath99b,marciano99a}.  
These parameters are $\tan\phi$ (the
ratio of wall-Higgs vev to bulk-Higgs vev) and
$R^{-1}\equiv M_c$ (the compactification scale).  We shall see below that
strong correlations exist between allowed values of $\tan\phi$,
$M_c$, and $m_H$ once we require that the 5DSM be consistent with all 
measurements.

%%%%%%%%%%%%%%%%%%%%%%%%%%%%%%%
\subsection{Global fits with physical observables}

The procedure for carrying out a global fit is the same for the HDSM as it
is for the SM:
\begin{itemize}
\item[1] Construct the full bare Lagrangian of the theory, 
     ${\cal L}(g_0,m_0, \psi_0,\ldots)$.
\item[2] Split the bare parameters and bare fields into renormalized 
    quantities and counterterms, 
    ${\cal L}(g+\delta g,m+\delta m,\psi+\delta Z_\psi \psi/2, \ldots)$.
\item[3]      Decide on a renormalization scheme (MS-bar, on-shell, etc.) that
      sets the values of the counterterms (e.g., set to a loop correction
     at a particular scale). For tree-level calculations, it is most
     convenient to set the counterterms to zero.
\item[4] Calculate all observables using the renormalized Lagrangian.  From the
   previous steps the result will be finite and depend only on
   renormalized couplings ${\cal O}_i(g,m,\ldots)$.  
\item[5] Perform a constrained global fit to see if there is a set of
     renormalized couplings $g,m,\ldots$ that allows
    ${\cal O}_i(g,m,\ldots )={\cal O}_i^{\rm expt}$ to within 
    experimental uncertainty.
\end{itemize}
In some cases a model can be completely ruled out by the above procedure,
whereas in other cases like the SM and the 5DSM, the model can work for
a limited range of parameter choices for the as-yet unknown parameters.

There are many observables that we wish to compare predictions with
experiment. Above, we specified the Lagrangian and renormalized
parameters that enable us to carry out this program.  In this section
we write down, analytically, the calculated observables at leading order
in an expansion of $m_W^2/M_c^2$.  We expand in $m_W^2/M_c^2$
({\it i.e.}, $V$) since we know that we recover the good SM fit to data 
as $M_c\gg m_W$.  The physical vector boson masses are to leading
order in $m_W^2/M_c^2$,
\bea
m_W^2 & = &\frac{1}{2}g^2 v^2\, [1-s_\phi^4 V], \\
m_Z^2 & = &\frac{1}{2}\frac{g^2}{\cos^2\theta} v^2\, [1-s_\phi^4 V/\cos^2\theta],
\eea
where we define
\beq
\label{Veq}
V\equiv 2\frac{m_W^2}{M_c^2}\sum_{n=1}^{\infty} \frac{1}{n^2} =\frac{\pi^2}{3}
   \frac{m_W^2}{M_c^2},
\eeq
and $g_2\equiv g$, $g_1\equiv g/\tan\theta$, and 
$v^2\equiv \langle \varphi_{\rm wall}\rangle^2+
       \langle \varphi_{\rm bulk}\rangle^2$.
The last equality in Eq.~\ref{Veq}
is valid only if $\ppar =4$ (one extra spatial dimension).
It is also convenient to define a charge-current and neutral-current 
interaction coupling with the lightest $W$ and $Z$ mass eigenvalues,
\bea
\frac{g_W}{\sqrt{2}}J_\mu^{\rm CC}W^\mu +{\rm h.c.} ~~~~ & \Longrightarrow &
   ~~~~\frac{g_W}{\sqrt{2}} = \frac{g}{\sqrt{2}}[1-s_\phi^2 V] \\
\frac{g_Z}{\cos\theta}J_\mu^{\rm NC}Z^\mu +{\rm h.c.} ~~~~ & \Longrightarrow &
   ~~~~\frac{g_Z}{\cos\theta} = 
      \frac{g}{\cos\theta}[1-s_\phi^2 V/\cos^2\theta ].
\eea

We can now express more easily other observables in terms of
the physical vector boson masses $m_W$ and $m_Z$ and the definitions
$g_W$ and $g_Z$ provided above.  For example, 
\beq
G_F(\mu~{\rm decay}) =  \frac{\sqrt{2}g_W^2}{8m_W^2}[1+V],
\eeq
\beq
\Gamma(Z\to f\bar f) =  \frac{N_c m_Z}{12\pi}
 \left( \frac{g_Z}{2\cos\theta}\right)^2 
  \left[ v_f^2 + a_f^2 \right], 
\eeq
\beq
A_f =  \frac{2v_fa_f}{v_f^2+a_f^2},
\eeq
\bea
Q_W & =& \frac{1}{m^2_Z}\left\{ \frac{g^2 (1-s^2_\phi V/\cos^2\theta )^2}
 {\cos^2\theta} + \frac{g^2V}{\cos^4\theta}\right\} 
      \nonumber\\
  & & ~~ \times a_e \left[ v_u (2Z+N)+v_d(2N+Z)\right] , 
\eea
\bea
R & = & \frac{\sigma^\nu_{\rm NC} -\sigma^{\bar \nu}_{\rm NC}}
 {\sigma^\nu_{\rm CC} -\sigma^{\bar \nu}_{\rm CC}} =
 \left[ \frac{g^2_Z}{(1-\sin^2\theta )m_Z^2}
          +\frac{g^2 V}{(1-\sin^2\theta)^2m_Z^2}\right] \nonumber \\
 & & ~~\times \left[ \frac{g^2_W}{m^2_W}+\frac{g^2V}{m^2_W}\right]^{-1} 
 \left( \frac{1}{2}-\sin^2\theta\right) ,
\eea
\beq
\sin^2\theta^{\rm eff}_W = x + \frac{x(1-x)}{1-2x}V
   \left[ c^4_\phi-\frac{s^4_\phi}{1-x}\right],
\eeq
\beq
m^2_W=m^2_Z(1-x)\left\{ 1+V\left[ 1-2s^2_\phi-
       \frac{c_\phi^4(1-x)-s^4_\phi}{1-2x}\right]\right\},
\eeq
where $Q_W$ is a measure of atomic parity violation, $x$ is the solution
to the equation
\beq
\label{xeq}
x(1-x)=\frac{\pi\alpha}{\sqrt{2}G_F m^2_Z},
\eeq
and,
\beq
v_f\equiv T_{3f}-2Q_f\sin^2\theta ~~~~{\rm and}~~~~a_f\equiv T_{3f}.
\eeq

All observables depend explicitly or  
implicitly on $V$ since renormalized
parameters such as $g$ and $\sin^2\theta$ are merely intermediate book-keeping
devices in the pursuit of expressing observables in terms of other
observables.  The best-fit values from data of the renormalized parameters
will depend, for example, on how much $V$ affects $G_F$.  

There are also important loop
corrections to these observables.  We assume that the loop corrections
involving KK excitations are higher order corrections compared to
loop corrections from zero-mode particles (``SM states'')
and tree-level KK interactions with
the zero modes.  Furthermore, on the $Z$-pole we ignore the tree-level
contribution
of exchanged $\gamma^{(n)}$ and $Z^{(n)}$ KK excitations to the
total background (off-resonant) rate.  
This is justified since $Z$-pole scattering
does not interfere with off-shell background processes. Although 
ordinary photon exchange subtraction is necessary when translating
raw $Z$-pole data into $Z$ decay rates, the high KK mass assumption 
($M_c\gg m_Z$) renders additional subtractions unnecessary.  The loop
corrections involving light zero-mode states 
are performed numerically with the aid of {\tt ZFITTER}~\cite{bardin94a}.

%%%%%%%%%%%%%%%%%%%%%%%%%%%%%%%%%%%%%%%%%%%%%%%%%
\subsection{Numerical results}

We have numerically carried out a $\chi^2$ global fit 
analysis of experimental data to
the HDSM.  The observables which we include in this analysis are,
\bea
\Gamma_{l^+l^-} & = & 83.90\pm 0.10 \mev~~\cite{dilella} \\
m_W & = & 80.410 \pm 0.044\gev~~\cite{dilella} \\
\sin^2\theta^{\rm eff}_W & = & 0.23157 \pm 0.00018~~\cite{dilella} \\
R_b & = & 0.21680 \pm 0.00073 ~~\cite{dilella}\\
R_c & = & 0.1694 \pm 0.0038 ~~\cite{dilella}\\
Q_W & = & -72.06 \pm 0.28 \pm 0.34~~\cite{bennett99a} \\
\sin^2\theta_W^{\nu N} & = & 0.2254 \pm 0.0021 ~~\cite{mcfarland98a}. 
%m_t & = & 173.8\pm 5.0~~\cite{partridge98} .
\eea
In this fit we have held fixed
$m_Z=91.1867\gev$~\cite{lepewwg98},
$G_F= 1.16637\times 10^{-5}\gev^{-2}$~\cite{stuart99},
$\alpha_s(m_Z)=0.119$~\cite{lepewwg98}, $m_t=173.8\gev$~\cite{dilella}, and
$\alpha_{\rm QED}(m_Z)= 1/128.933$~\cite{davier98}.

We assume that one physical Higgs 
scalar boson is present in the spectrum which interacts with
the fermions and gauge bosons like a SM Higgs boson.  The other physical 
Higgs degrees of freedom either do not exist or have interactions
decoupled from the zero modes of the gauge bosons and the fermions. 
This is analogous to the MSSM, where one Higgs boson acts like a SM
Higgs boson and the rest decouple, being irrelevant for precision
measurement analyses.

Our procedure, then, is to choose a Higgs boson mass and vary
$V$ to see how the predictions change for the observables.  We wish
to minimize the $\chi^2$ function defined as,
\beq
\chi^2\equiv \sum_i 
  \frac{({\cal O}^{\rm theory}_i(g,m_h,V,\ldots)-{\cal O}_i^{\rm expt})^2}
   {(\Delta {\cal O}^{\rm expt}_i)^2}.
\eeq
We also define $\Delta\chi^2= \chi^2-\chi^2_{\rm min}$.

In Fig.~\ref{Vlimit2} we plot $\chi^2$ with $\tan\phi=0$
and for differing choices of $m_h$.  In the SM, the 95\% C.L. upper
bound on the Higgs mass is $260\gev$~\cite{lepewwg98}.  In this
plot the $\chi^2$ value for $V=0$ (decoupled extra dimensions) and
$m_h=260\gev$ is $\chi^2=22.1$. We then allow $V$ to vary from zero
and $m_h$ to vary, and define the allowed region of parameter space
to be that which has $\chi^2<22.1$.  
From Fig.~\ref{Vlimit2} we can see that the light Higgs boson
is favored for $V=0$, just as the well-known SM results indicate.  
Furthermore, as we increase the Higgs mass the best fit value of
$V$ drifts more and more into the $V<0$ region.
Within the context of the 5DSM,
negative values of $V$ are not physical.  Increasing the value
of $V$, or, equivalently in the 5DSM, lowering the compactification
scale $M_c$, we see that the electroweak precision data fit only
gets worse for any value of $m_h$.  The largest value of $V$ with
$\chi^2<22.1$ is $V=0.0015$.  Therefore, the limit on $V$ in
this theory is $V<0.0015$ which is equivalent to $M_c>3.8\tev$ in
the 5DSM.
%%%%%%%%%%%%%%%%%%%%%%%%%%%%%%%%
\jfig{Vlimit2}{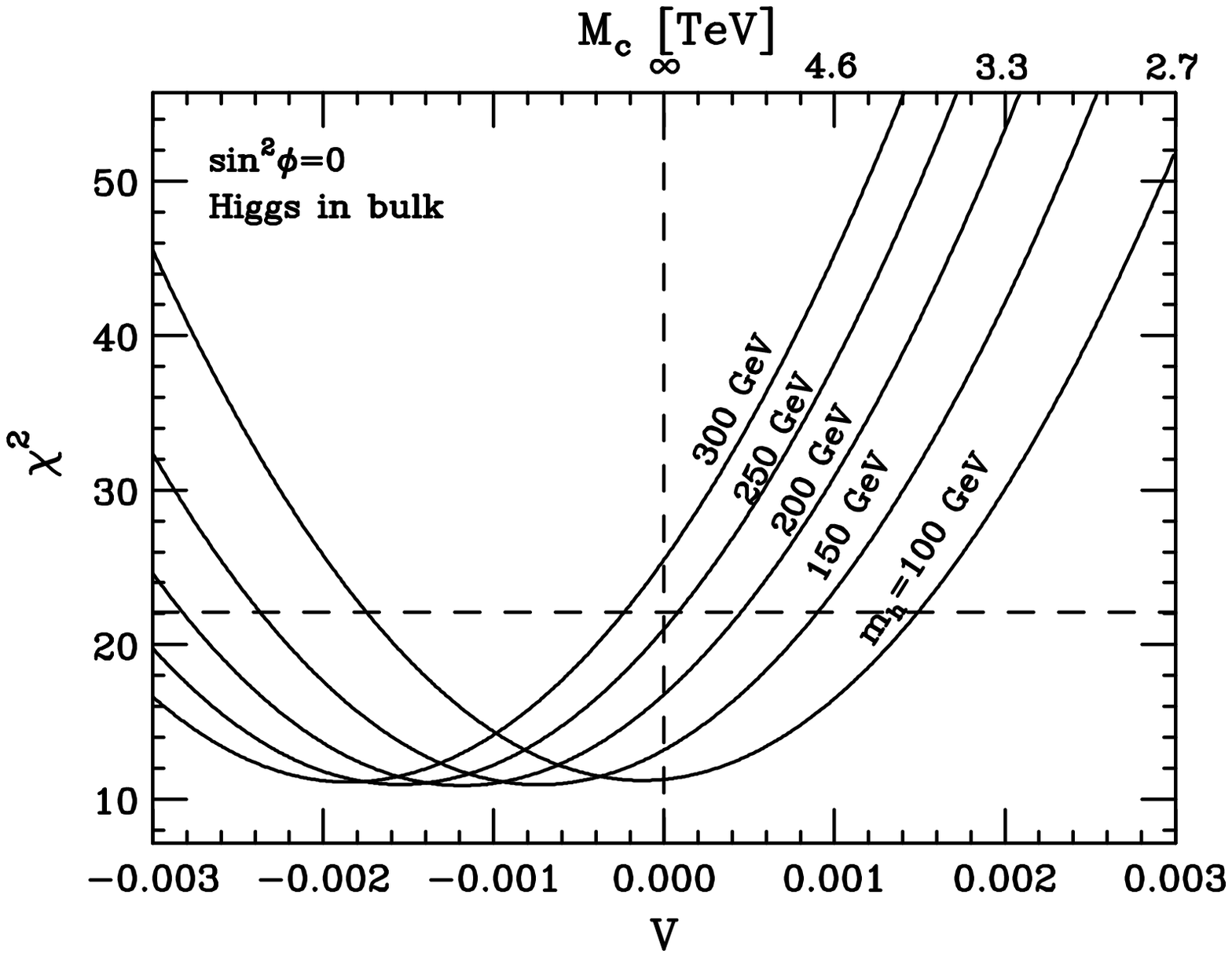}{$\chi^2$ to precision electroweak
data with $\tan\phi =0$ (Higgs bosons in bulk only).  The horizontal
dashed line is for $\chi^2_{\rm SM,max}=22.1$, 
which reproduces the $\chi^2$ for
$V=0$ and $m_h=260\gev$, which is currently the 95% C.L. confidence level for
the Higgs mass in the SM.  Requiring $\chi^2<\chi^2_{\rm SM,max}$
implies the limit $V<0.0015$, which translates to 
$M_c>3.8\tev$ in the 5DSM.}
%%%%%%%%%%%%%%%%%%%%%%%%%%%%%%%%%%%%%%%%%%%%%%%%%%%%%%%%%%%%

For $\tan\phi=\infty$, meaning the only Higgs boson(s) associated with
EWSB is on the wall, we find the opposite behavior.  In Fig.~\ref{Vlimit3}
we plot the $\chi^2$ for various choices of $m_h$, with
$\tan\phi=\infty$ and with $V$ varying.  In this case, 
the fit remains good as $V$ increases and $m_h$ increases.
(Note, the $V=0$ slice
is equivalent to the $V=0$ slice of Fig.~\ref{Vlimit2}.) 
A similar
relaxing of the SM Higgs boson mass limit from precision data
has been demonstrated in other contexts~\cite{hall99a,barbieri99a}.
Now, all the minima of the
$\chi^2$ fits are in the $V>0$ physical region.  For $m_h>260\gev$,
a non-zero value of $V$ is required to be present in the theory in order to
provide an acceptable fit to the data. 
As demonstrated in Fig.~\ref{Vlimit3}, the Higgs mass could be heavy and as
high as $500\gev$ and still have $\chi^2<22.1$ as long as $V\simeq 0.0016$.
That is, KK excitations of gauge bosons must
substantially affect precision electroweak predictions in order
to obtain a good fit with $m_h>260\gev$.
If the Higgs mass gets above $500\gev$ then there is no longer a 
choice of $V$ for which $\chi^2<22.1$.  Limits on $V$ can also be obtained
by finding its maximum value with $\chi^2<22.1$.  This value
is $V<0.002$ which translates to $M_c>3.3\tev$ in the 5DSM.
%%%%%%%%%%%%%%%%%%%%%%%%%%%%%%%%%%%%%%%%%%%
\jfig{Vlimit3}{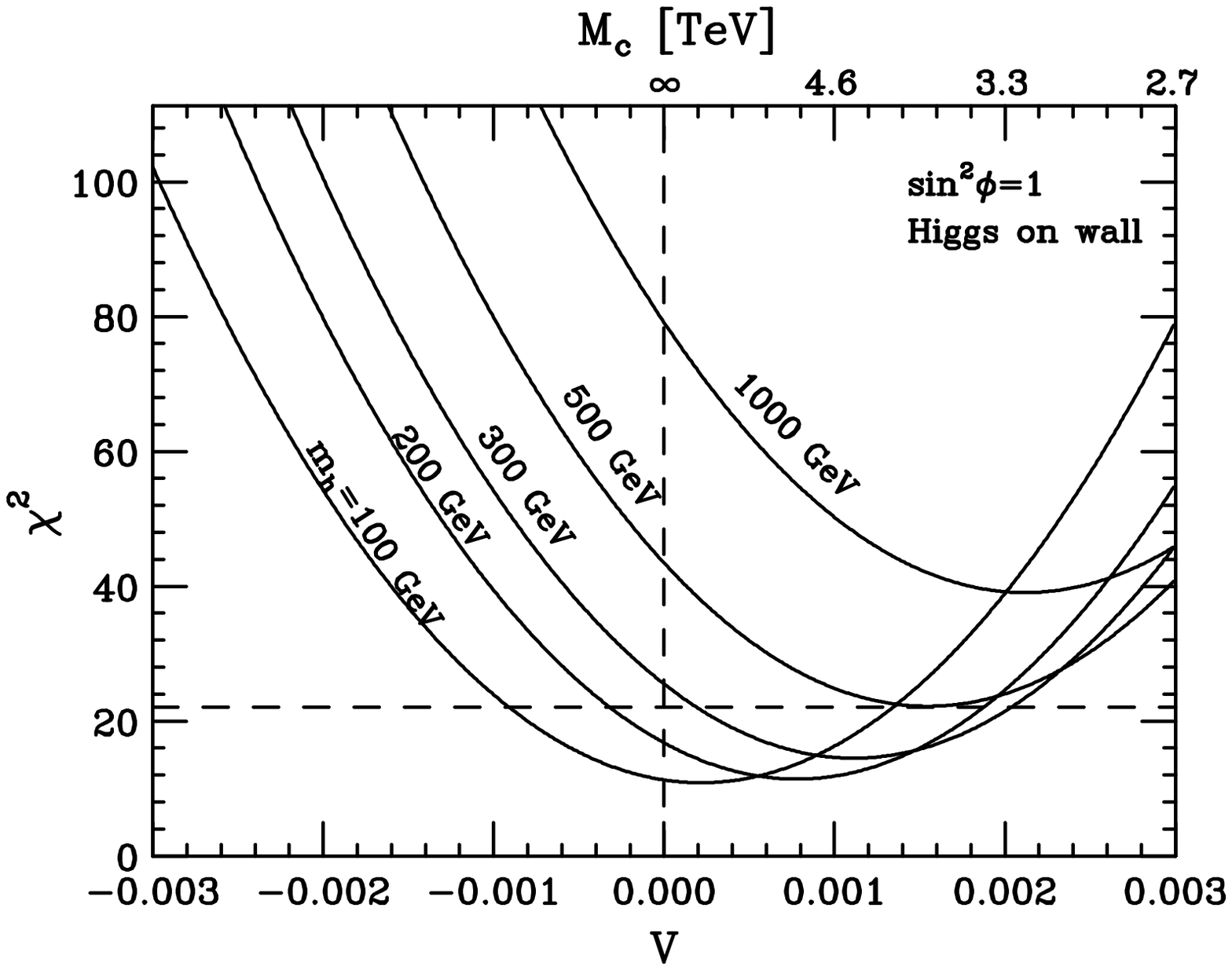}{$\chi^2$ to precision electroweak
data with $\tan\phi =\infty$ (Higgs bosons on 3-brane wall only).
The horizontal
dashed line is for $\chi^2_{\rm SM,max}=22.1$, 
which reproduces the $\chi^2$ for
$V=0$ and $m_h=260\gev$, which is currently the 95% C.L. confidence level for
the Higgs mass in the SM.  Requiring $\chi^2<\chi^2_{\rm SM,max}$
implies the limit $V<0.002$, which translates to 
$M_c>3.3\tev$ in the 5DSM.  Furthermore, values of $m_h$ as high as
$500\gev$ are allowed as long as $V>0$.}
%%%%%%%%%%%%%%%%%%%%%%%%%%%%%%%%%%%%%%%%%%%%%%%%%%

The reason why large $m_h$ is compatible with precision data can be
seen most clearly by inspecting the behavior of $\sin^2\theta^{\rm eff}_W$
and $m_W$ in the limit of $\sin^2\phi=1.0$.  For $V>0$ the value
of $\sin^2\theta^{\rm eff}_W$ decreases and $m_W$ increases, precisely
what lowering the Higgs boson mass
would do to the predictions.
In this case, the low-mass Higgs boson is not needed if
$V$ is sufficiently high.

We next ask what the 95\% C.L. range is for $V$ given a fixed $m_h$.
This question differs slightly from the previous question, in that
we are no longer asking how good of a fit a particular value of
$m_h$ is, but rather what deviations of $V$ would be tolerated
if $m_h$ were given to us from another source (direct experiment, ``by god'', 
etc.).  For this we must analyze the $\Delta \chi^2$ distribution,
which is defined to be $\chi^2-\chi^2_{\rm min}$, where $\chi^2_{\rm min}$
is the lowest value of $\chi^2$ for a fixed $m_h$ but variable $V$.
Then, the 95\% range of $V$ is determined by requiring $\Delta\chi^2<(1.96)^2$.
The case where a negative
$V$ provides the best fit must be handled by following the 
Feldman-Cousins prescription~\cite{feldman98}.
In Table~1 we show these ranges of $V$ for a given $m_h$ and $\sin^2\phi$.
The blank spaces in the table mean that there is no range
of $V$ allowed in the physical region, and the parenthesis mean that
there is no choice of $V$ for that particular $m_h$ and $\sin^2\phi$
which gives $\chi^2<22.1$.  Therefore, any non-blank entry 
without parenthesis means that the corresponding Higgs boson mass is
not ruled out for that given choice of $\sin^2\phi$ and at least
one value of $V$.
\begin{table}
\label{table1}
\centering
\begin{tabular}{cccccc}
\hline\hline
$m_h$ [GeV] & $\sin^2\phi=0$ & $\sin^2\phi=0.25$ & $\sin^2\phi=0.50$ & $\sin^2\phi=0.75$ & $\sin^2\phi=1.0$ \\
\hline
100 & $<0.83$ & $<1.56$ & $<2.25$ & $<1.42$ & $<0.89$ \\
150 & 0.40 & 0.86 & 2.22 & 1.83 & 1.21 \\
200 & 0.25 & 0.53 & 2.15 & $0.15-2.09$ & $0.20-1.45$ \\
250 & 0.20 & 0.40 & 2.08 & $0.33-2.30$ & $0.34-1.63$ \\
300 &      & 0.33 & (2.01) & $0.46-2.48$ & $0.44-1.77$ \\
350 &      &      & (1.96) & $0.57-2.67$ & $0.54-1.87$ \\
400 &      &      & (1.90) & ($0.63-2.75$) & $0.69-2.03$ \\
500 &      &      & (1.81) & ($0.85-2.97$) & $0.88-2.22$ \\
600 &      &      & (1.71) & ($1.02-3.14$) & ($1.03-2.39$) \\
700 &      &      & (1.65) & ($1.16-3.28$) & ($1.16-2.52$) \\
800 &      &      & (1.59) & ($1.29-3.41$) & ($1.25 - 2.61$) \\
900 &      &      & (1.54) & ($1.39-3.51$) & ($1.34-2.70$) \\
1000 &     &      & (1.50) & ($1.49-3.61$) & ($1.45-2.79$) \\
\hline\hline
\end{tabular}
\caption{{ 95\% C.L. ($\Delta\chi^2 < (1.96)^2$) 
allowed ranges of $V/10^{-3}$ for different values of 
$\sin^2\phi$ and $m_h$ in the global $\Delta \chi^2$
distribution to precision electroweak data.
Blank spaces in the table mean that the fit to the data is too poor 
to quote a bound 
in the physical region of $V>0$. Parenthesis means that there is no value
of $V$ such that $\chi^2<22.1$, implying that the corresponding Higgs
boson mass is not allowed from the global $\chi^2$.}}
\end{table}

%%%%%%%%%%%%%%%%%%%%%%%%%%%%%%%%%%%%%%%%%%%%%%%%%%%%%%%%%%%%
\section{Kaluza-Klein excitations at high-energy colliders}
\bigskip

The results of the previous section were obtained by comparing
precision measurements of electroweak observables to the theoretical
predictions of the HDSM.  These results were mainly derived from
how the zero-mode vector bosons interact with the KK excitations
and how the KK excitations of the $W$ and $Z$ directly affect observables
with characteristic energy below $m_Z$ ($\mu$ decay, $\nu N$ scattering, 
etc.).
In this section, we
estimate the sensitivity of KK excitations to $e^+e^-$ scattering
at high-energy colliders above $m_Z$.  
This will involve operators induced by higher
modes of the $W/Z/\gamma$ gauge bosons and also on-shell production
of KK excitations of the SM gauge bosons.

%%%%%%%%%%%%%%%%%%%%%%%%%%%%%%%%%%%%%%%%%%%%%%%5
\subsection{Indirect searches at $e^+e^-$ colliders}

The observables we wish to study arise from amplitudes induced by the
KK excitations of gauge bosons,
\beq
{\cal A}(e^+e^-\to f\bar f)=\sum_{n=0}^{\infty} A(e^+e^-\to \gamma^{(n)}/Z^{(n)}\to f\bar f).
\eeq
As $M_c$ gets larger the excited modes of $\gamma^{(n)}$ and $Z^{(n)}$ obtain heavier
and heavier mass and have minimal impact on the overall scattering amplitude.
The amplitude contributions from the excited modes can be analyzed
effectively by integrating out the heavy modes and constructing operators
which take into account all their effects.  For example, integrating out the
higher modes of the photon yields an operator of the form
\beq
{\cal O}_{\gamma^{(n)}} = \sum_{\vec n}^\infty 
     \left( \frac{ -2g^2\sin^2\theta Q_eQ_f}{\vec n^2M_c^2}\right)
               [\bar e\gamma_\mu e][\bar f \gamma^\mu f]
     = -g^2\sin^2\theta Q_eQ_f \frac{V}{m_W^2}
        [\bar e\gamma_\mu e][\bar f \gamma^\mu f].
\eeq
Similar operators arise from integrating out 
$W^{(n)}$ and $Z^{(n)}$,
\bea
{\cal O}_{W^{(n)}} & = & \frac{-g^2V}{2m_W^2} 
       [\bar e\gamma_\mu P_L \nu][\bar f\gamma^\mu P_L f'], \\
{\cal O}_{Z^{(n)}} & = & \frac{-g^2V}{4\cos^2\theta m_W^2}
     [\bar e\gamma_\mu (v_e-a_e\gamma_5)e][ \bar f\gamma^\mu (v_f-a_f\gamma_5)
     f].
\eea

Limits can be set on $V$ from the effect of these operators on
the total rates and polarization asymmetries of $e^+e^-\to f\bar f$ for
all accessible fermions (see~\cite{rizzo96} for a discussion
of the observables).  In Fig.~\ref{lep2} we plot the search reach
of $V$ versus integrated luminosity for
$\sqrt{s}=195\gev$. To construct this plot we have included initial
state radiation with  a 10 degrees polar angle cut on the photons.
The $b$ and $c$ quark tagging efficiencies are taken to be 35\% and 20\%
respectively. We also assume that the KK states decay only into SM particles.
The conclusions may be weakened if these KK excitations were to decay
some fraction of the time into superpartners. 
With over $1\xfb^{-1}$ one can either detect or rule
out $V>2.4\times 10^{-3}$ (or $M_c<3.1\tev$ for the 5DSM).
%%%%%%%%%%%%%%%%%%%%%%%%%%%%%%%%%
\jfig{lep2}{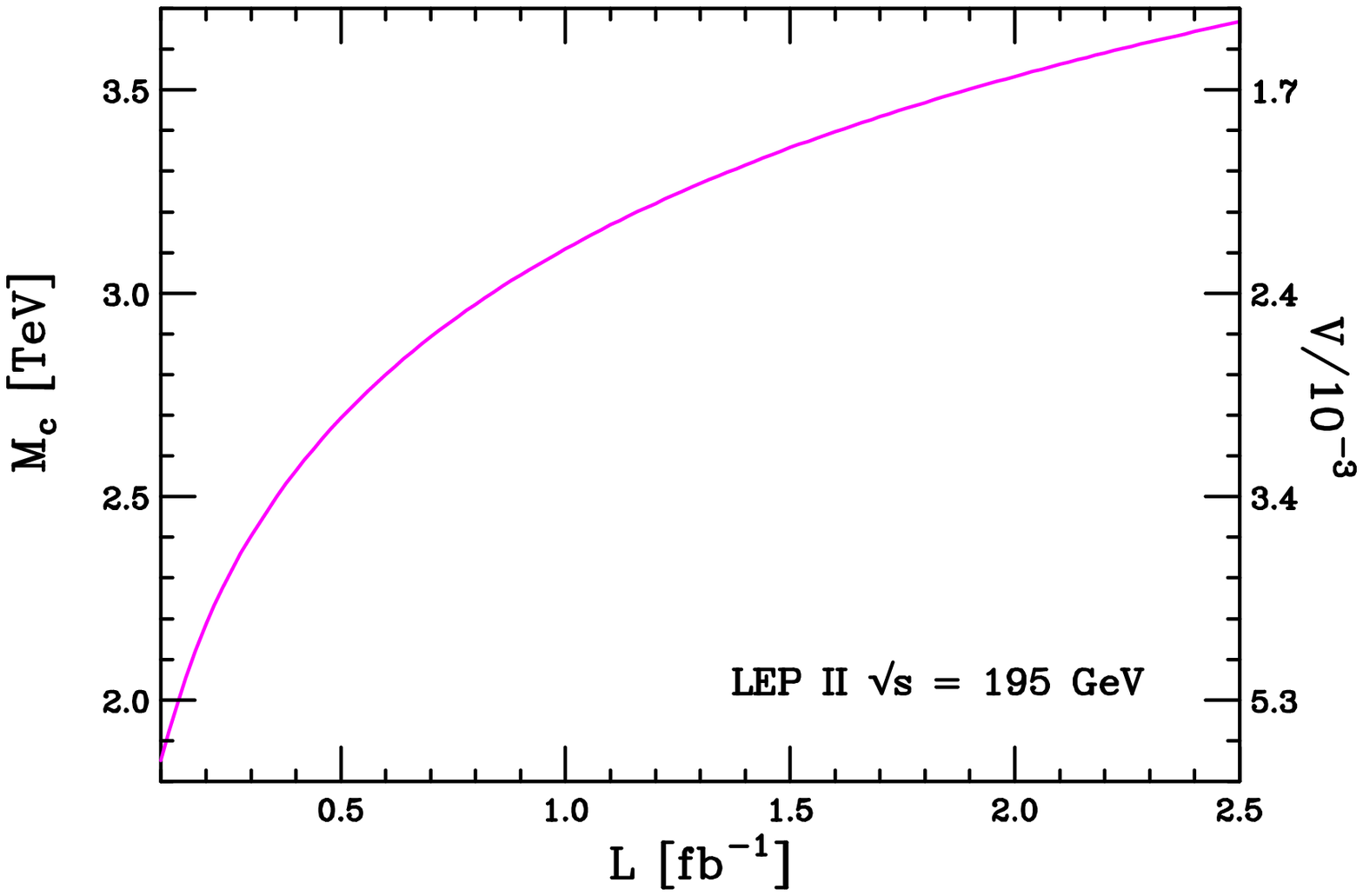}{Search reach in $V$ at LEP2 running
at $\sqrt{s}=195\gev$ as a function of integrated luminosity.  
For the 5DSM the value of $V$ can
be related to $M_c$, which is shown on the left vertical axis. }
%%%%%%%%%%%%%%%%%%%%%%%%%%%%%%%%%%%%%

The same analysis can be applied at the NLC.  However, here we are
able to add the top quark to the list of final states.  Furthermore, 
we can include
polarization asymmetries at the NLC, and we can utilize observables
associated with $\tau$ lepton polarization.  At the NLC we 
assume
that $b$, $t$ and $c$ quark identification efficiencies are 
$60\%$ and the efficiency for measuring tau polarization is 
50\%~\cite{nlc96,rizzo96}.

The sensitivity to $V$ at the NLC is substantial.  In Fig.~\ref{nlc}
we plot the search reach for $V$ versus integrated luminosity
for $\sqrt{s}=500\gev$, $1000\gev$ and $1500\gev$ 
$e^+e^-$ colliders.  With more than $500\xfb^{-1}$ the reach
is $V>12.5\times 10^{-5}$, $V>4.0\times 10^{-5}$, 
and $V>2.2\times 10^{-5}$ for the three ascending center-of-mass
energies.  In 5DSM, we can translate these reaches in $V$ into reaches
of $M_c$ and find  $13\tev$, $23\tev$ and $31\tev$ respectively.
These are significantly greater than for a typical GUT-inspired
$Z'$~\cite{rizzo96} due to 1) the larger couplings ({\it i.e.,}
$\sqrt{2}$ enhancement), 2) the tower contribution, and 3)
both $Z^{(n)}$ and $\gamma^{(n)}$ contribute.
%%%%%%%%%%%%%%%%%%%%%%%%%%%%%%%%%%%%%%%%%%%%%%%%%%%%%%%%%%%5
\jfig{nlc}{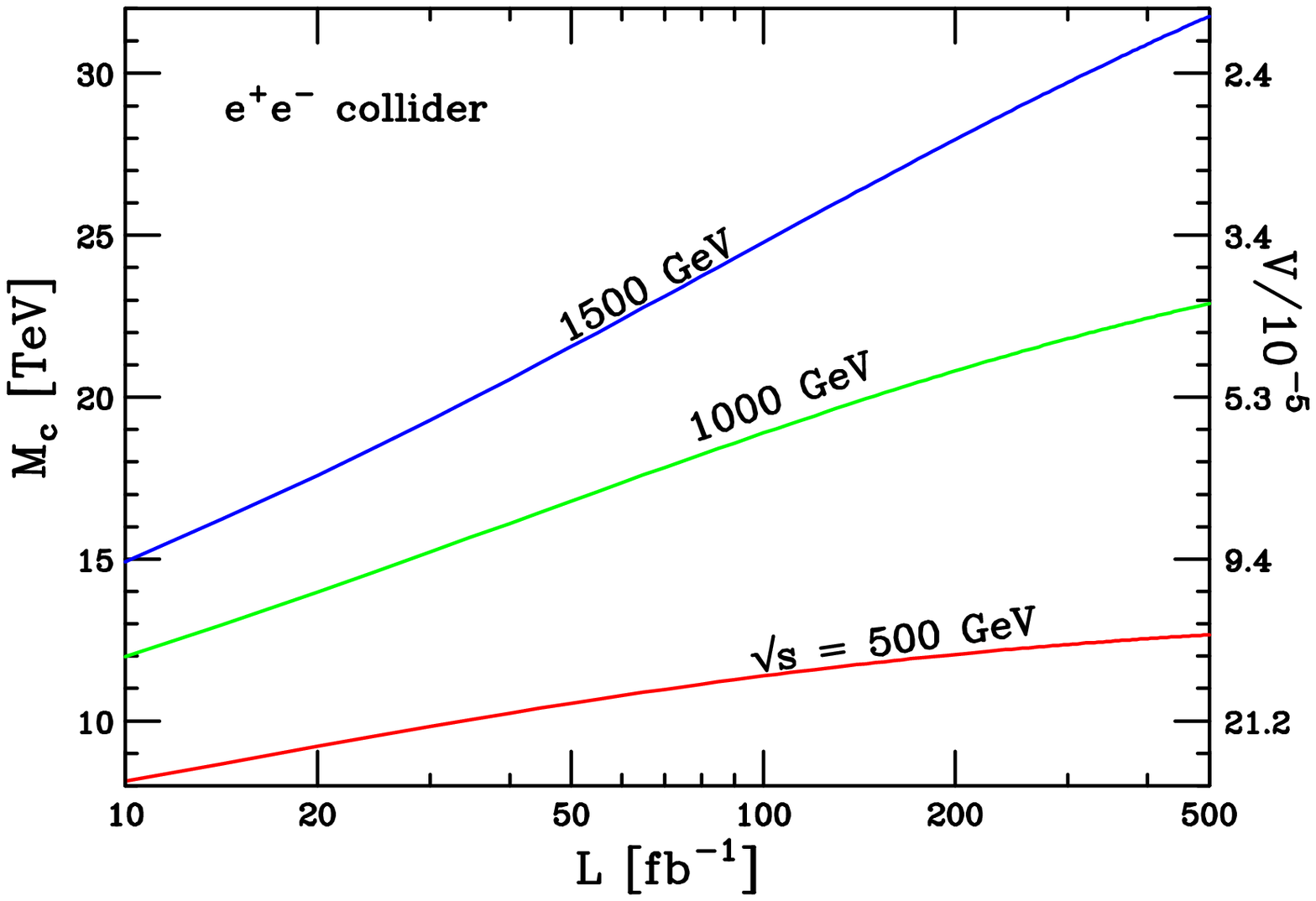}{Search reach in $V$ at NLC running
at $\sqrt{s}=500\gev$, $1000\gev$ and $1500\gev$ as a function
of integrated luminosity.  
For the 5DSM the value of $V$ can
be related to $M_c$, which is shown on the left vertical axis.}
%%%%%%%%%%%%%%%%%%%%%%%%%%%%%%%%%%%%%%%%%%%%%%%%%%%%%%%%%%%%%%%

%%%%%%%%%%%%%%%%%%%%%%%%%%%%%%%%%%%%%%%%%%%%%%%%%%%%%
\subsection{Direct searches at hadron colliders}
\bigskip

One can also look for direct production of the KK states at
hadron colliders. We do not consider the capabilities of
indirect, off-shell contributions of the KK excitations to hadron collider
observables, since $Z'$ studies indicate that the resonant production
searches are more probing. The neutral current mode of producing final
state lepton pairs in $Z^{(n)}/\gamma^{(n)}$ mediated Drell-Yan
processes is the most useful mode to search for evidence of
extra dimensions at the Tevatron and LHC.

The scattering processes $f\bar f\to l^+ l^-$ 
through intermediate KK excitations $Z^{(n)}$ and $\gamma^{(n)}$
leads to peaks in the invariant mass spectrum at high-energies.
The couplings of $Z^{(n)}$ and $\gamma^{(n)}$ to fermions is
the same as their corresponding zero-mode couplings, except for
an overall enhancement of $\sqrt{2}$.  
%Furthermore, the multiplicity
%of states at each excitation level depends on the number of extra
%dimensions.  For example, the number of $Z$ excitations at the
%first level with mass $M_c$ is equal to the number of extra dimensions.
%Therefore, the more extra dimensions there are the larger the
%cross-section through $f\bar f\to Z^{(1)}/\gamma^{(1)}\to l^+l^-$.
In our analysis we estimate the sensitivity to $M_c$ from only the
first excited state.  Incorporating other excited states into the search
would yield a slightly higher sensitivity than what we present here.

The search strategy is based on leptonic final states in the narrow
width approximation.
For $Z^{(1)}$ first excited state with high mass, 
the search is for a narrow high-energy
dilepton invariant mass excess.  
For $W^{(1)}$ at high mass, the search is for a
high-energy lepton plus large missing energy.  For a given luminosity
the cuts are chosen such that no Standard Model background events are
expected, and a signal is declared if there are more than 10 events presents
given the same cuts.  The strategy is summarized in ref.~\cite{rizzo96}.

We can now present search capabilities for 1 extra dimension, the 5DSM, 
at the Tevatron and LHC.
At the Tevatron with
$\sqrt{s}=2\tev$ and integrated luminosity of $2\xfb^{-1}$
($20\xfb^{-1}$) we find that $M_c$ up to $1.20\tev$ ($1.36\tev$)
can be probed in the combined neutral channels mediated by
$\gamma^{(1)}$ and $Z^{(1)}$.  
At the LHC, in this same channel with $\sqrt{s}=14\tev$ and $100\xfb^{-1}$
integrated luminosity, we find a reach of up to $5.9\tev$.  Searches
for the $W^{(1)}$ mode at the Tevatron allow discovery at $1.11\tev$
and $1.34\tev$ with $2\xfb^{-1}$ and
$20\xfb^{-1}$ respectively.  The $W^{(1)}$ can be discovered at the
LHC with $100\xfb^{-1}$ if its mass is less than $6.35\tev$. 

One could also look for enhancements in the dijet production at high 
invariant mass from KK excitations of the gluons.  The procedure we employ
here is similar to that used to constrain resonant production of squarks in
theories with light gluinos~\cite{hewett96}.  We have extrapolated the CDF and
D0 data to higher energies and luminosity, and find a reach capability
of $M_c\lsim 700\gev$, which is somewhat lower than the reach capability
from $Z^{(1)}/\gamma^{(1)}$ and $W^{(1)}$ induced processes.  At the LHC
we estimate the reach of the first gluon excited level
to be below $5\tev$, although the precise number depends sensitively on
the dijet energy resolution.

The search reach increases with more extra dimensions because there
are more copies of the first KK excitation gauge bosons.  For example,
with one extra dimension presented above, there is only one copy of
$Z^{(1)}$, $\gamma^{(1)}$ and $W^{(1)}$.  With $d$ extra dimensions there
are $d$ copies of these bosons, making a higher production rate of 
final state leptons for the same $M_c$.  Also, with only one extra dimension
the next KK excitation level is at $2M_c$ where only one copy of the
gauge bosons reside, whereas with more than one
extra dimension the next KK level is lower, $\sqrt{2}M_c$, where there
may be many copies of the SM gauge bosons.  Therefore, as the number of extra
dimensions increases, it appears to become more important to consider 
the higher KK levels to get an accurate estimate of the maximum search
reach.  However, as discussed in section~\ref{applicability}, the 
naive effective
field theory description of KK excitations cannot be correct, especially for
more than one extra dimension, and the couplings of the higher KK states
must necessarily be suppressed in a model dependent way.  For this reason,
we have focussed only on the first excited state.

%%%%%%%%%%%%%%%%%%%%%%%%%%%%%%%%%%%%%%%%%%%%%%%%%%%%%%%%%%%%5
\section{Conclusions}
\bigskip

In conclusion, the Standard Model originating from more than four extra
dimensions is just as good of a description of nature as the 4d Standard
Model.  The difference is in the allowed physical parameters that have not
yet been detected.  For example, in the ordinary 4DSM, the Higgs boson
mass must be less than about $260\gev$ in order for the precision
electroweak data to match the data well.  This is not the case in the 
5DSM, where much larger masses (up to $500\gev$)
for the Higgs boson are allowed as long as the
Higgs boson is confined to the wall and KK excitations of the
gauge bosons are rather light.

Table~2 contains a summary of many of the results presented
in the text.  All results have been translated into bounds or sensitivity
on the compactification scale in the 5DSM, where one can see that 
current and future colliders will
be able to probe well into the TeV region.  This is especially relevant
for the solution to the hierarchy problem, which we view as the strongest
motivations for this scenario.  If low-scale compactification
theories do have some relevance for electroweak symmetry breaking and
the hierarchy problem, it is then at the TeV scale that we expect evidence for
them to show up.  This is directly analogous to expectations for finding
supersymmetry, since low-scale supersymmetry also solves the hierarchy
problem.  The scale of $M_c$ can then be thought of in the same way
as the scale of superpartner masses and the ratio $m_W^2/M_c^2$ is one
measure of how fine-tuned the electroweak potential is.  It is for these
reasons that we are optimistic that low-scale, sub-Planckian compactifications 
are more likely at lower
scales near $m_W$ than at higher, inaccessible scales.
\begin{table}
\label{summary}
\centering
\begin{tabular}{lc}
\hline\hline
Experiment & $M_c$ reach \\
\hline
PEW with Higgs in bulk & $3.8\tev$ \\
PEW with Higgs on wall & $3.3\tev$ \\
LEPII with $\sqrt{s}=195\gev$ and $L=1\xfb^{-1}$ &  $3.1\tev$ \\
Tevatron with $\sqrt{s}=2\tev$ and $L=2\xfb^{-1}$ & $1.1\tev$ \\
Tevatron with $\sqrt{s}=2\tev$ and $L=20\xfb^{-1}$ & $1.3\tev$ \\
LHC with $\sqrt{s}=14\tev$ and $L=100\xfb^{-1}$ & $6.3\tev$ \\
NLC with $\sqrt{s}=500\gev$ and $L=500\xfb^{-1}$ & $13\tev$ \\
NLC with $\sqrt{s}=1000\gev$ and $L=500\xfb^{-1}$ & $23\tev$ \\
NLC with $\sqrt{s}=1500\gev$ and $L=500\xfb^{-1}$ & $31\tev$ \\
\hline\hline
\end{tabular}
\caption{{ Summary of search reaches of the compactification scale
in the 5DSM with different experiments.  PEW stands for precision
electroweak data accumulated at LEP, SLD, NuTeV, etc.  The hadron collider
numbers are for direct production sensitivity of the first KK excited
states of the gauge bosons, and the high-energy $e^+e^-$ collider
numbers are for inferred limits from deviations in fermion final state
observables below on-shell threshold.}}
\end{table}

\bigskip
\noindent
{\it Acknowledgements:} We thank F.~del~Aguila, T.~Gherghetta, G.~Giudice,
J.~Hewett, M.~Masip and A.~Pomarol for helpful conversations. TGR thanks
the CERN Theory Division, where part of this was done, for its
hospitality.

%%%%%%%%%%%%%%%%%%%%%%%%%%%%%%%%%%%%%%%%%%%%%%%%%%%%%%%%%%%%%%%%%%%%%%%%

\end{document}